\documentclass[aps,prapplied,twocolumn,showpacs,superscriptaddress,longbibliography]{revtex4-1}
\usepackage{graphicx}
\usepackage{gensymb}
\usepackage{amsmath}
\usepackage{enumerate}
\usepackage{xcolor}
\newcommand{\upperRomannumeral}[1]{\uppercase\expandafter{\romannumeral#1}}
\newcommand{\lowerromannumeral}[1]{\romannumeral#1\relax}

\begin{document}
\title{Valley-Hall In-Plane Edge States as Building Blocks for Elastodynamic Logic Circuits}
\author{Jihong Ma}
\affiliation{Department of Civil, Environmental, and Geo- Engineering, University of Minnesota, Minneapolis, MN 55455, USA}
\author{Kai Sun}
\affiliation{Department of Physics, University of Michigan, Ann Arbor, MI 48109, USA}
\author{Stefano Gonella}
\email{sgonella@umn.edu}
\affiliation{Department of Civil, Environmental, and Geo- Engineering, University of Minnesota, Minneapolis, MN 55455, USA}

\begin{abstract}
In this work, we investigate theoretically and demonstrate experimentally the existence of valley-Hall edge states in the in-plane dynamics of honeycomb lattices with bi-valued strut thickness. We exploit these states to achieve non-trivial waveguiding of optical modes that is immune to backscattering from sharp corners. We also present how different types of interfaces can be combined into multi-branch junctions to form complex waveguide paths and realize a variety of structural logic designs with unconventional wave transport capabilities. We illustrate this potential with two applications. The first is a direction-selective energy-splitting waveguide tree featuring a pronounced asymmetric wave transport behavior. The second is an internal waveguide loop along which the energy can be temporarily trapped and periodically released, effectively working as a signal delayer. The modal complexity of in-plane elasticity has important consequences on the regime of manifestation of the edge states, as the availability of viable total bandgaps is shifted to higher frequencies compared to the out-of-plane counterpart problem. It also poses additional experimental challenges, associated with proper acquisition and deciphering of the in-plane modes, the solution of which requires a systematic use of in-plane laser vibrometry.
\end{abstract}

\maketitle
\section{\label{sec:Intro}Introduction}
Acousto-elastic metamaterials and phononic crystals are architected structures that are artificially designed to control and manipulate mechanical wave propagation. Some of their most well known properties are mechanical filtering and wave directionality \cite{kushwaha1993acoustic, liu2000locally, ruzzene2003wave, gonella2008analysis}, waveguiding (conventional and sub-wavelength) \cite{khelif2004guiding, celli2015manipulating, lemoult2016soda} and energy trapping \cite{jiao2018intermodal,zhao2015broadband, colombi2014sub}. In recent years, topological insulators (TI) endowed with topologically-protected edge states (TPES) have paved the way to new strategies for wave manipulation in quantum and electronic systems \cite{hasan2010colloquium,moore2010birth,qi2010quantum,chang2013experimental,hsieh2008topological,kane2005z}. Inspired by these condensed matter problems, researchers have been able to adapt the working principle of TI to acoustic and elastic systems to achieve analogous wave control capabilities in both the static \cite{kane2014topological,paulose2015topological,rocklin2017transformable,rocklin2016mechanical,bilal2017intrinsically} and dynamic regime \cite{prodan2017dynamical,nash2015topological,wang2015topological,khanikaev2015topologically,souslov2017topological,susstrunk2015observation,mousavi2015topologically,he2016acoustic,Ma2018edge}.

One of the first quantum phenomena to inspire a mechanical analog was the quantum Hall effect (QHE), which allows one-way, non-reciprocal wave propagation by creating chiral edge states that are robust against defects and disorders \cite{nash2015topological, wang2015topological,khanikaev2015topologically,souslov2017topological}. The QHE can be realized by breaking time-reversal symmetry through the application of an external (e.g. magnetic) field \cite{zhang2005experimental} that results in the lifting of a degenerate Dirac cone in the band diagram of lattice and the opening of a non-trivial bandgap. In a corresponding mechanical system, it is possible to resort to external active controls, such as spinning rotors \cite{nash2015topological,wang2015topological} or circulating fluids \cite{khanikaev2015topologically,souslov2017topological}, to achieve analogous unidirectionally propagating edge states. However,  unlike in electronic systems, the complexity required to apply external mechanical fields partially hinders the practicality of QHE for applications at the device level. In light of these limitations, researchers started exploiting the mechanical analog of the quantum spin Hall effect (QSHE), for which time-reversal symmetry breaking is not required. The QSHE can be realized by manipulating sublattice configurations to create mechanical pseudo-spins and pseudo-spin-dependent effective edges, ultimately achieving topologically protected helical edge modes that are also immune to back-scattering \cite {susstrunk2015observation,mousavi2015topologically,he2016acoustic}. While enjoying the inherent advantages of being fully passive, QSHE requires carefully engineered configurations in order to obtain the doubly-degenerate Dirac cones whose lifting is essential to establish helical edge wave conditions. As a result, its practical applicability is also met with difficulties. Mechanical systems involving topologically-protected floppy edge modes \cite{kane2014topological,paulose2015topological,rocklin2017transformable,rocklin2016mechanical,Ma2018edge} have also been proposed to obtain asymmetric wave propagation. Although these systems can be easily manufactured and indeed achieve asymmetric wave transport without requiring any active control, the asymmetric behavior only occurs at frequencies lower than the folding region of the acoustic modes \cite{Ma2018edge}.

The most recent efforts in this line of developments have been directed towards creating mechanical analogs of the quantum valley Hall effect (QVHE), whose only requirement is the breaking of space-inversion symmetry (SIS), which can be achieved in a relatively agile fashion in many acoustic and mechanical systems. In acoustics, this effect has been demonstrated for sonic crystals obtained from arrays of rod-like scatterers \cite{lu2016valley,lu2017observation}. In elastodynamics, the concept was first realized in honeycombs with masses added at lattice nodes \cite{pal2017edge,vila2017observation}, and then extended to lattices with curved struts \cite{liu2018tunable,liu2018experimental} and thin plates endowed with arrays of surface masses \cite{chen2017topological, chaunsali2018experimental, zhu2018design,makwana2018geometrically,makwana2018designing}. In hexagonal lattices with a $C_{6v}$ symmetry, Dirac points, or valleys are observed at the high-symmetry corners of the  Brillouin zone. SIS breaking, which preserves $C_{3v}$ symmetry, results in bandgap opening and in the lifting of the cones. This leaves behind two adjacent valleys,  $\boldsymbol{K}$ and  $\boldsymbol{K}'$, that are largely separated in the reciprocal space and feature opposite values of the locally-defined topological index known as valley Chern number. The Chern number dichotomy between the valleys implies low modal compatibility between phonons associated with the $\boldsymbol{K}$ and $\boldsymbol{K}'$ points (which can be interpreted as waves traveling in opposite directions in the lattice), which ultimately manifests as a low degree of back-scattering. A complete and lucid description of the working principle of QVHE is given in Ref. \cite{liu2018experimental}. The SIS breaking requirement is relatively simple to achieve through a mechanical modulation of the unit cell parameters, making this approach much more amenable to manufacturing and therefore applicable to elastic systems. Most efforts on QVHE to date (with the notable exception of one theoretical work on in-plane elasticity in discrete systems \cite{chen2018study}) have been devoted to the analysis of the out-of-plane flexural (antisymmetric Lamb wave) response of thin elastic plates.

In this work, we shift our attention to the in-plane dynamics of thin lattices in plane-stress conditions. Realizing viable QVHE conditions in in-plane systems features some additional challenges compared to their out-plane counterparts. In out-of-plane problems, a total bandgap can be easily obtained at low frequencies \cite{pal2017edge,vila2017observation,lu2016valley,lu2017observation}. This allows a straightforward establishment of edge waves that are not contaminated by any of the bulk modes (as a side note, it is worth noting that edge modes have also been successfully achieved in systems and frequency regimes where SIS breaking only induces partial bandgap opening \cite{liu2018tunable, zhu2018design}; this result can be explained by invoking the low modal compatibility and the large difference in density of states between the edge and bulk modes at the frequencies of interests, which makes the establishment of edge modes more favorable). In contrast, in-plane mechanics are characterized by additional modal complexity associated with the co-existence of longitudinal and shear modes even at low frequencies. As a result, identifying total bandgaps that naturally allow for spectrally-isolated edge modes inevitably requires exploring the high-frequency optical regime.

Our structure of choice is a modified hexagonal lattice consisting of beam-like struts with bi-valued thickness. This configuration is characterized by extreme geometric simplicity, which makes it easy to manufacture using conventional cutting or printing techniques. While this structure features multiple bandgaps that are amenable to QVHE, here we focus our attention on the high-frequency regime, where a wide total bandgap is available. Under these conditions, we study numerically and demonstrate experimentally the establishment of non-trivial waveguides and their robustness against back-scattering from sharp corners. Moreover, we leverage this platform to realize complicated paths involving multi-branch junctions and different types of domain walls to achieve a variety of special energy channeling and routing effects. As a first example, we design a direction-selective energy-splitting waveguide tree that allows one-way wave propagation at high frequencies. We also design a triangular waveguide loop that works as a signal delayer: here a portion of the energy injected in the system is temporarily trapped in the loop, and then released periodically at an output boundary with programmable delays.

\begin{figure} [ht]
	\centering
	\includegraphics[scale=0.55]{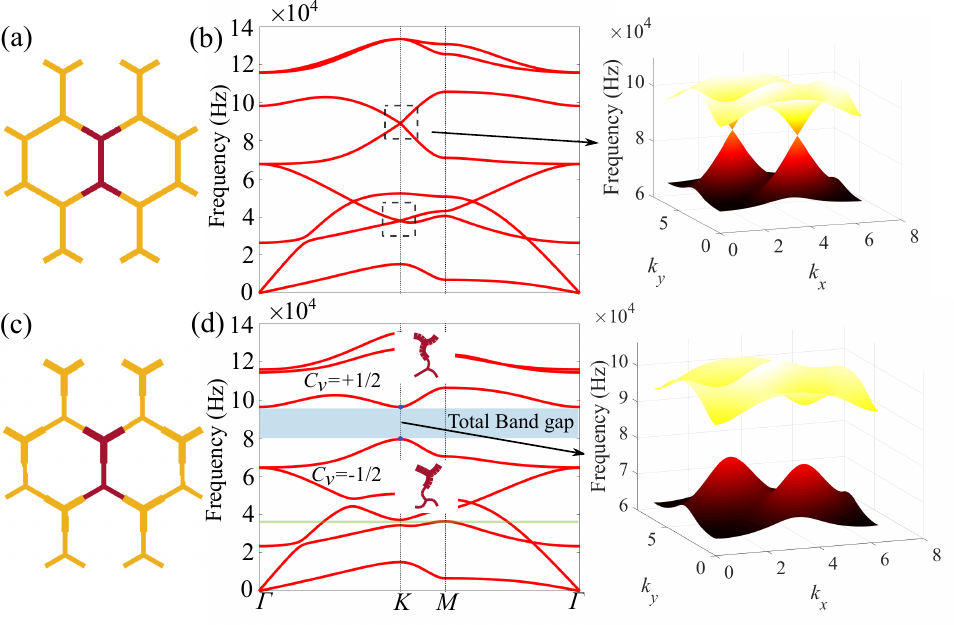} 
	\caption{(a) Unit cell (maroon) of a regular hexagonal lattice with uniform strut thickness, and (b) its corresponding band diagram. The crossing points in the dashed squares are Dirac points. The surface plot on the right provides a 3D visualization of the Dirac cones at the $\boldsymbol{K}$ point. (c) Unit cell (maroon) of a hexagonal lattice with modulated strut thickness. The thickness of the thick (thin) parts is increased (decreased) by 20\% with respect to the reference value in (a). (d) Band diagram for the bi-valued hexagonal lattice. The green and blue shaded areas denote partial and total band gaps, respectively. The two inserted displaced cells represent the eigen-displacements at the $\boldsymbol{K}$ points for the 5th and the 6th bands.}
	\label{unitcell}
\end{figure}

\section{Topologically-Protected Edge Modes}
\label{sec:theory}
To obtain a lattice with a $C_{3v}$ symmetry after breaking the SIS, we start with a regular hexagonal lattice with a $C_{6v}$ symmetry made of 2-cm-long struts with a 7.5:1 in-plane aspect ratio, Fig.~\ref{unitcell} (a). The geometric and material properties are chosen to match those of the specimen that we intend to test experimentally, which is manufactured via water jet cutting from a sheet of acrylonitrile butadiene styrene (ABS): length of the struts $l=2$ cm, out-of-plane sheet thickness $d=3.2$ mm, Young's modulus, $E=2.14$ GPa, Poisson's ratio $\nu=0.35$, density $\rho=1040$ $ \, \textrm{kg}/\textrm{m}^3$. The phonon dispersion relation, obtained via Bloch unit cell analysis conducted via finite element analysis (FEA), is shown in Fig. \ref{unitcell} (b). The struts are modeled as shear-deformable beams according to Timoshenko beam theory to expedite the calculation  with respect to a 2D elasticity model, without compromising the accuracy. In the frequency range of interest (below $1.0\times10^{5}$ Hz), dictated by the bandwidth of our actuation equipment, we identify two Dirac points at the $\boldsymbol{K}$ point. A detailed calculation of the Berry phase confirming the Dirac-like nature of these points, is reported in Section I of the Supplemental Material (SM) \cite{SM}. 

To break the SIS, we modulate the strut thickness so that the aspect ratio of the upper (lower) half of the unit cell is increased (decreased) by 20\%, as shown in Fig.~\ref{unitcell} (c). The asymmetry of the lattice opens the Dirac cones and creates band gaps of $1.684\times10^{4}$ Hz and $2.720\times10^{3}$ Hz for the upper and lower cones, respectively, see Fig. \ref{unitcell} (d). We decide to focus our analysis on the upper gap, which is wider and more tractable. To describe the topology of this gap, we calculate the valley Chern number $C_{v}$ at the valley $\boldsymbol{K}$ ($\boldsymbol{K}'$) point, and obtain the value $+(-) 0.48$, which approximately tends to $+(-)\frac{1}{2}$ of classical QVHE theory.

\begin{figure} [h]
	\centering
	\includegraphics[scale=0.26]{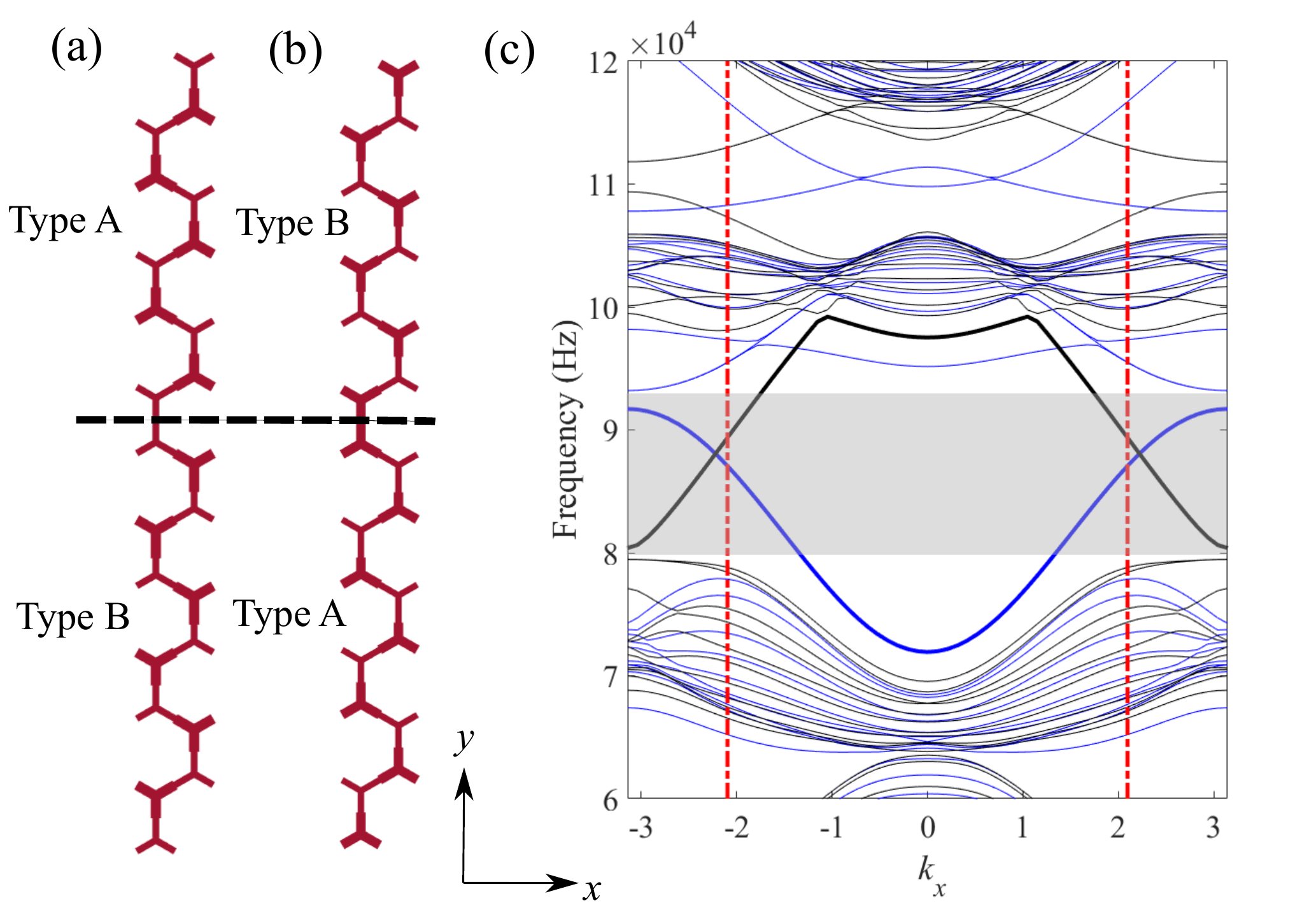} 
	\caption{(a) and (b): Supercells featuring the two types of non-trivial interfaces along the zigzag directions, with (a) thin and (b) thick interfaces connecting the two lattice types. (c) Band diagram of the supercells. Blue and black curves refer to the thin and thick interfaces shown in (a) and (b), respectively. The thick curves denote the interface modes. The gray area marks the non-trivial band gap. The dashed red lines indicate the location of $\boldsymbol{K}'$ (left) and $\boldsymbol{K}$ (right) points.}
	\label{sc}
\end{figure}

We now proceed to consider a lattice consisting of two domains (termed lattice type A and B) separated by an interface. The cells above the interface have thin (thick) struts in the top half of the cell and thick (thin) struts in the bottom half, while those below the interface are mirror symmetric, as shown in Fig. \ref{sc} (a) [(b)]. To confirm the existence of interface states, we conduct a $\it{supercell}$ analysis. Each supercell consists of a finite strip of 12 cells: the top six cells belong to type A (B), while the bottom six to type B (A), as shown in Fig. \ref{sc} (a) [(b)]. In this way, the lattice resulting from the repetition of the supercell along the $x$-axis displays a horizontal interface along the zigzag direction. The asymmetry of the unit cell allows two possible types of interface: we name an interface where two thin struts meet a $\it{thin}$ interface, Fig. \ref{sc} (a), and the other type $\it{thick}$ interface, Fig. \ref{sc} (b). We apply 1D Bloch conditions along the $x$-axis while leaving the top and bottom boundaries free to mimic a finite-width, infinitely long horizontal strip. From the supercell band diagrams obtained for the two different interfaces, superimposed in Fig. ~\ref{sc} (c), we detect the emergence of two new bands within the bandgap. The mode shapes calculated at the valley  $\boldsymbol{K}$ point ($k_x=2\pi/3$) on these new bands, reported in Section I of the SM \cite{SM}, indicate that the emerging bands are indeed interface modes.

\section{\label{sec:exp}Experimental evidence of topologically-protected in-plane waveguiding}

An important feature of a TPES is that it allows waves to propagate along sharp corners with negligible back-scattering. Effective rationales for this phenomenon have been proposed in the literature in the context of flexural waves. In Section II of the SM \cite{SM}, we recall and extend these considerations, providing additional insight in the process that is germane to our specific problem of in-plane wave propagation.

\begin{figure} [h]
	\includegraphics[scale=0.17]{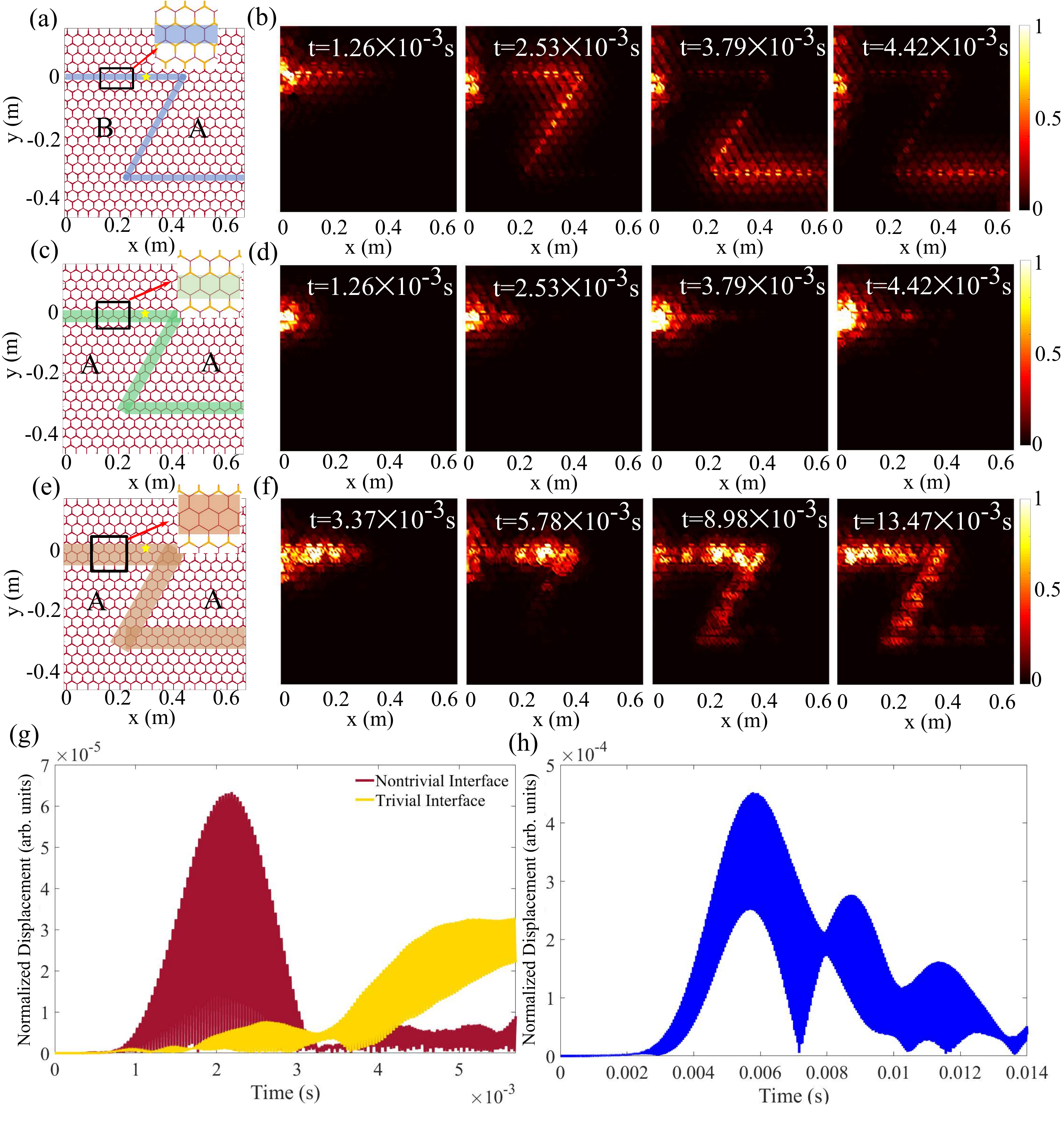} 
	\caption{Full-scale simulations of lattices with (a) nontrivial (blue shade), (c) narrow trivial (green shade), and (e) wide trivial (red shade) interfaces. The two regions, A and B, separated by the nontrivial interface in (a) present a mirror symmetry about the interface. For trivial interfaces in (c) and (e) the two regions have the same lattice type. (b), (d) and (f) are the wavefields for the (a) nontrivial, (c) narrow and (e) wide trivial interfaces, respectively. (g) Magnitude of displacement at the  locations marked by a star along the narrow trivial (yellow) and nontrivial (maroon) interfaces. (h) Magnitude of displacement along the wide trivial (blue) interface. The displacements in (g) and (h) are normalized by the largest measured values.}
	\label{interface}
\end{figure}

Here, we focus our attention on verifying this phenomenon, both numerically and experimentally, for in-plane phonons in our lattice of choice. To document the robustness against back-scattering, we model a Z-shape thin interface with two $60^\circ$ corners, shown in Fig. \ref{interface} (a). The interface splits the lattice into two sub-domains with mirror symmetry about the interface. The dynamic response is computed via full-scale dynamic FEA. A burst excitation with frequency $\omega=8.707\times10^4$ Hz with 35 cycles is applied in the form of a horizontal in-plane force at left edge of the interface. As is evident from Fig. \ref{interface} (b), the wave indeed propagates along the interface without significant back-scattering until it reaches the right boundary. The back-scattering-immune behavior can also be confirmed by plotting the displacement time history of an arbitrary point located along the interface, as shown in Fig. \ref{interface} (g). The presence of a single dominant peak, associated with the incoming wave, indicates that no appreciable reflections are generated at the corners. Since the excitation frequency is within the bulk band gap, no leakage into the bulk is observed. 

An alternative strategy to realize a similarly-shaped waveguide is by constructing a trivial interface. This can be done starting from a uniform lattice (say type A) and replacing the cells featuring thickness modulation with regular hexagonal cells along the interface. Let us consider two trivial Z-shaped domain walls with interface widths spanning over one and two cells, as shown in Fig. \ref{interface} (c) and (e), respectively. Interestingly, when the trivial interface is one cell in width, as in Fig.~\ref{interface} (c), the wave remains mostly localized around the excitation point, with little energy propagating along the interface, as shown in Fig. \ref{interface} (d). Therefore, a trivial interface with one-cell thickness is not sufficient to effectively close the band gap and establish efficient waveguide transport conditions. By sufficiently widening the interface, as in the case of Fig. \ref{interface} (e), involving a two-cell interface, waves can eventually propagate along the trivial waveguide, as shown in Fig.~\ref{interface} (f). However, strong back-scattering is observed at corners, significantly weakening the signal transmitted and rendering this a very inefficient waveguide. This is best seen by looking at Fig. \ref{interface} (h), which again shows the time history of the response sampled at a point along the first segment of the waveguide. A second peak at around $9\times10^{-3}$ s is clear evidence of a strong reflected wave. In conclusion, only the topologically-protected interface guarantees efficient and scattering-free waveguiding.

\begin{figure} [th]
	\includegraphics[scale=0.155]{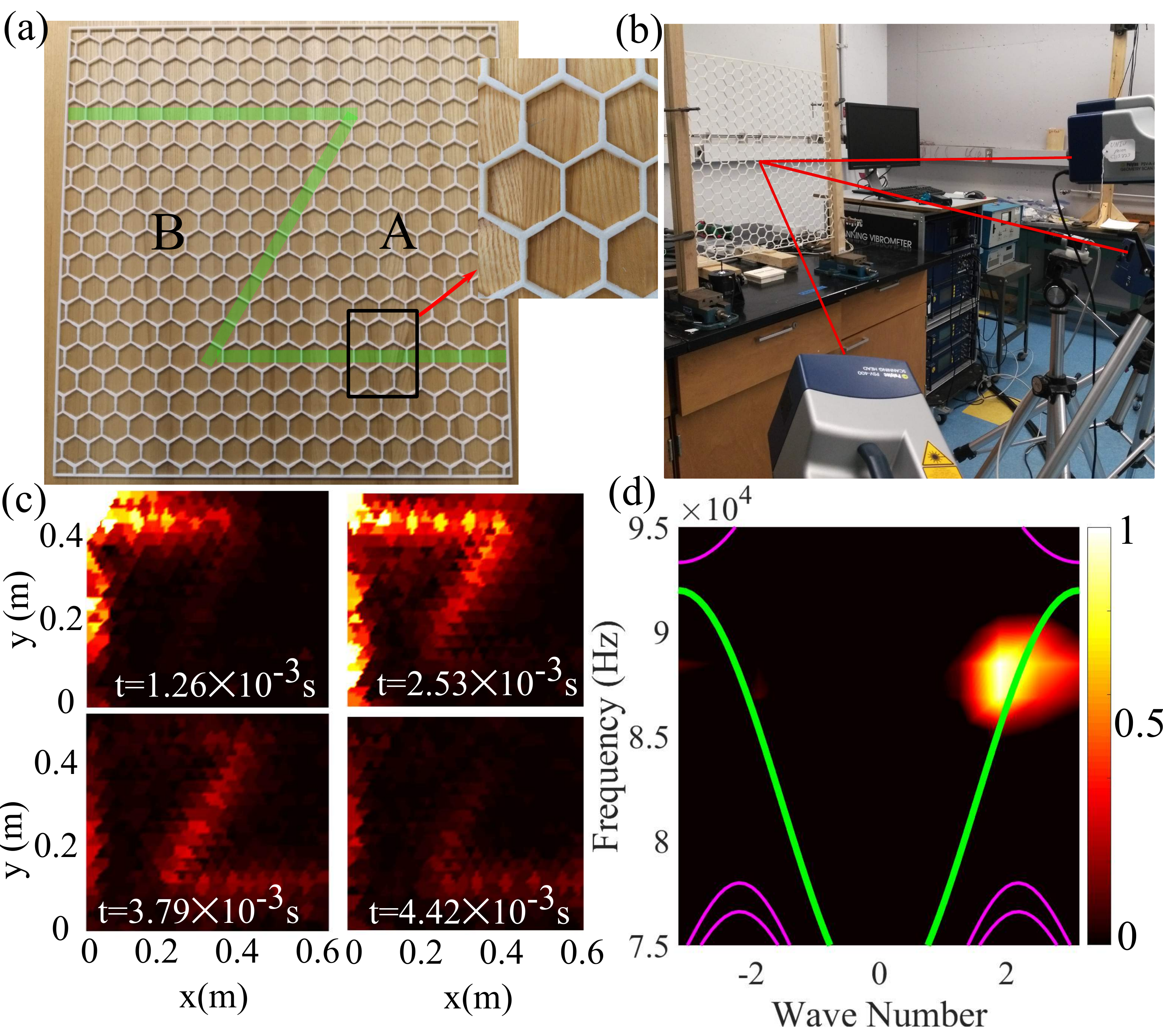} 
	\caption{(a) Experimental specimen cut from a thin ABS sheet. A Z-shape thin interface with two $60^\circ$ corners highlighted in green. (b) Experimental setup showing the 3D SLDV heads and the specimen rotated by $90^\circ$ counter-clockwise in order to facilitate excitation from the shaker. (c) Snapshots of the wavefield induced through burst excitation at $\omega=8.707 \times 10^4$ Hz. The point velocity magnitudes are normalized by the largest measured value. The magnitude is attenuated due to damping in the specimen. (d) DFT of the experimental data sampled along the interface, matching the interface mode (green) calculated from supercell analysis and revealing the absence of appreciable back-scattering. The magenta curves are bulk bands.}
	\label{experimentZ}
\end{figure}

We then move on to perform a series of experiments to confirm the topological attributes of the non-trivial waveguide. To acquire the in-plane wavefields, we scan the lattice with a 3D Scanning Laser Doppler Vibrometer (SLDV, Polytec PSV-400-3D) using one scan point per lattice node. The specimen is a water-jet-cut ABS lattice, comprising $34\times20$ unit cells with the same dimensions and parameters used in simulations, see Fig. \ref{experimentZ} (a). The green-shaded area denotes the Z-shape interface. For convenience of setup, we rotate the lattice by $90^\circ$ counter-clockwise and frame it using clamps at the two sides, leaving the top and bottom boundaries free. The excitation is prescribed as an in-plane normal point force applied on the boundary at the opening of the topological interface. The force is exerted using an electrodynamic shaker (Bruel \& Kjaer Type 4809, powered by a Bruel \& Kjaer Type 2718 amplifier) placed at the bottom of the lattice, as shown in Fig. \ref{experimentZ} (b). We prescribe a burst excitation at $\omega=8.707 \times 10^4$ Hz with 35 cycles in order to activate a very narrow frequency interval falling entirely on the topological edge mode branch within the bandgap, approximately at $k_x=2\pi/3$. From visual inspection of the wavefield, we observe that the wave indeed propagates along the interface without appreciable backscattering, see Fig. \ref{experimentZ} (c), consistent with the numerical prediction in Fig. \ref{interface} (b). Note that the magnitude is here attenuated along the propagation path due to unavoidable damping in the physical specimen. The robustness against back-scattering is eloquently captured by the spectral plane representation of Fig. \ref{experimentZ} (d), obtained by applying a discrete Fourier transform (DFT) to the experimental data sampled along the interface. Indeed, the spectral magnitude map presents a strong asymmetry between the $+k_x$ and $-k_x$ directions, suggesting negligible reflection from the corner. It can also be appreciated that the spectral response features excellent agreement with the edge mode branch obtained from supercell analysis. 

\section{\label{sec:app}Multi-interface Junctions for Energy Transport Manipulations }

In Section II of the SM \cite{SM}, we offer a rationale for how waves propagate around $60^\circ$ and $120^\circ$ corners with or without changes of interface type, and the conclusions align with what has been observed in out-of-plane counterpart problems \cite{pal2017edge,vila2017observation,liu2018tunable,liu2018experimental,chen2017topological, chaunsali2018experimental, zhu2018design,makwana2018geometrically,makwana2018designing}. These considerations can be fully leveraged for structural logic circuit designs with unconventional energy transport capabilities. Inspired by former theoretical studies of topological current splitters based on the valley Hall effect \cite{qiao2014current,kang2018pseudo}, we devote our attention to the wave manipulation and opportunities of multi-interface junctions in cellular metastructures. Recently, a study of topological junctions in phononic plates has documented the potential of junctions as energy splitters and filters for flexural (out-of-plane) waves, introducing the notion of tessellated mechanical domains as topological super networks \cite{makwana2018designing}. Here, we provide a systematic exploration of analog opportunities available in cellular configurations undergoing in-plane elastic wave propagation using a suite of FEA and laser vibrometry experiments. 

The objective is to exploit the energy splitting properties of topological junctions to achieve a series of new asymmetric wave transport and energy trapping effects, thus expanding the gallery of technological potentials of valleytronic concepts in mechanical systems. We start by noting that, according to our previous discussion, the lattice types on the two sides of a topologically-protected interface must present a mirror symmetry. Hence, to ensure that all the interfaces converging at a junction are topologically protected, it is necessary that the junction connects an \textit{even} number of interfaces. Here we illustrate this idea through two examples. 

\begin{figure} [th]
	\includegraphics[scale=0.21]{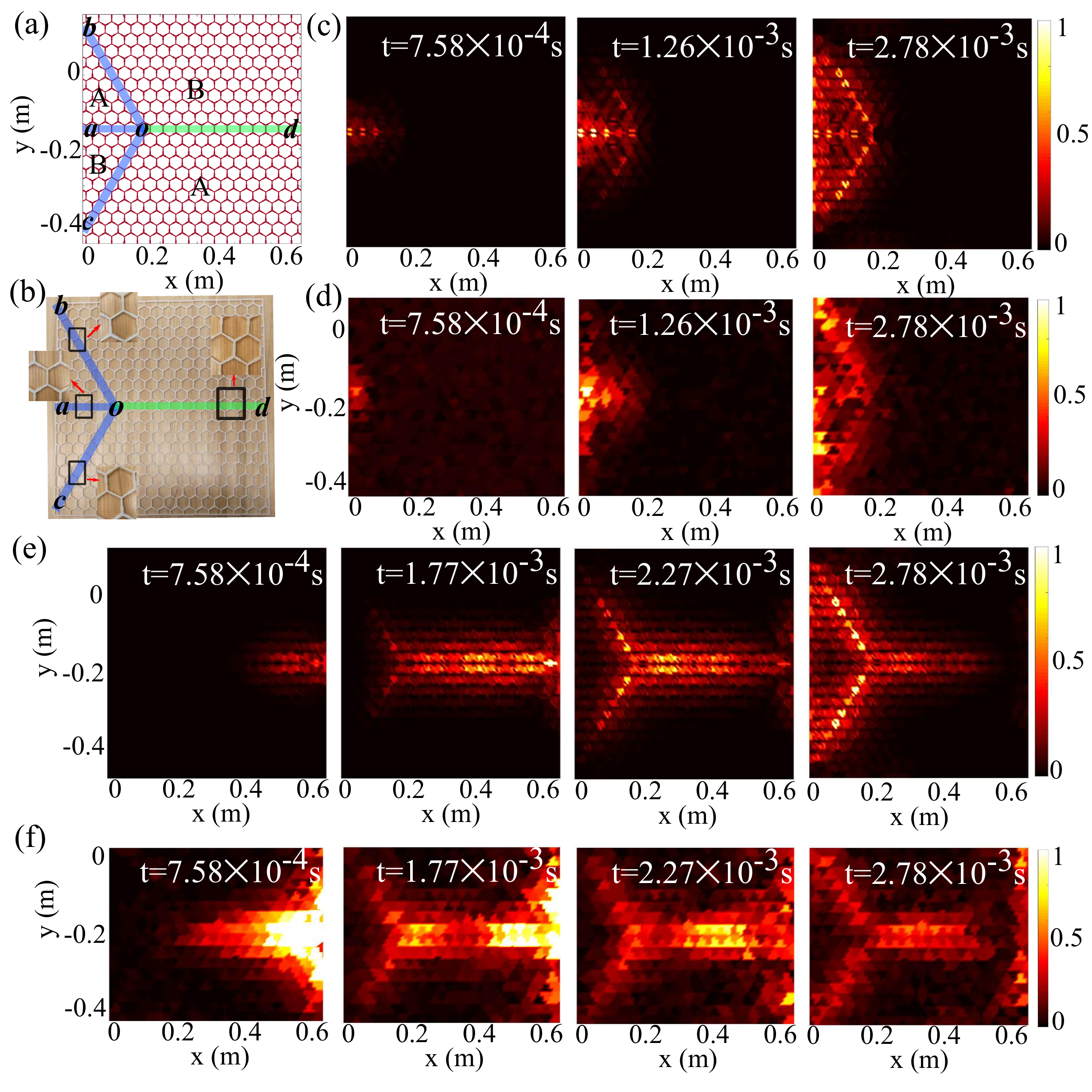} 
	\caption{(a) FEA model and (b) experimental specimen of a one-way elastic de-multiplexer. $a$-$d$ are four ports on the edges. Three thin interfaces, $ao$, $ob$, and $oc$, forming $60^\circ$ angles, are marked by blue shades. A thick interface $od$ connecting the right boundary to junction $o$ and forming a $120^\circ$ angle with two of the thin interfaces is denoted by a green shade. (c) and (d): Snapshots of wavefields obtained from (c) simulation and (d) experiment with burst excitation at $\omega=8.707 \times 10^4$ Hz applied at port $a$. (e) and (f): Snapshots of wavefields obtained from (e) simulation and (f) experiment with excitation at port $d$. The velocity magnitudes are normalized by the largest measured value. In the experimental case, the magnitude is attenuated due to damping.}
	\label{asymmetry}
\end{figure}

The first design is a direction-selective elastic energy splitter intended to allow strongly asymmetric wave transport at high frequencies. The FEA model and experimental specimen are shown in Fig. \ref{asymmetry} (a) and (b), respectively. Three ``thin'' and one ``thick'' interfaces (according to the terminology introduced before) are connected at point $o$, resulting in three ports on the left side of the lattice and one on the right. The angle between the three thin interfaces is $60^\circ$, while the one between the thin and thick interfaces is $120^\circ$. A burst excitation with carrier frequency falling inside the bandgap is applied at port $a$ on the left, and the simulated and experimentally measured wavefields are shown in Fig.~\ref{asymmetry} (c) and (d), respectively. It can be seen that the wave prefers to split and steer at the junction and propagate back along $ob$ and $oc$, eventually localizing at the left edge of the lattice without ever reaching the opposite edge. In contrast, the simulated and experimental wavefields obtained exciting at port $d$ on the right are shown in Fig.~\ref{asymmetry} (e) and (f), respectively. As we can see, the wave propagates along the thick interface from right to left and, at junction $o$, splits (with negligible reflections) into two transmitted signals traveling to ports $b$ and $c$. The characteristics of the interfaces coalescing at junction $o$ ensures that the wave propagates along the two thin interfaces $ob$ and $oc$ that form a $120^\circ$ angle with the thick interface $od$. As a result, transmission form right to left is achieved efficiently. Note that opposite group velocities between the thin $oa$ and the thick $od$ interfaces prevent the wave from propagating straight through the junction, despite the perfect alignment between $oa$ and $od$.

De facto, the lattice endowed with the internal multi-path junction behaves dynamically as a medium with effective asymmetrical wave transport characteristics. It works as an isolator for excitations prescribed at the left edge and as a conductor for excitations prescribed at the right edge. In this respect, it achieves a functionality that is a high-frequency analog of what has been recently shown experimentally for topological Maxwell structural lattices (with ligaments with finite bending stiffness), see Ref. \cite{Ma2018edge}. In such systems we observe a similar behavior whereby excitations prescribed at the floppy edge remain localized close to the excitation, while those prescribed at the stiff boundary are transmitted across the sample. Interestingly, despite the fact that the two effects are based on different mechanisms, they are both edge phenomena that can be seen as manifestation of the bulk-boundary correspondence, they both enjoy topological protection and they are both described by topological invariants (the valley Chern number here and the topological polarization vector in Ref. \cite{Ma2018edge}). From an application standpoint, the asymmetrical transport, more specifically, the emergence of an isolating edge results in high transmission efficiency, as the signal that has traveled across the sample is prevented from traveling back towards the input edge, even after scattering form the output edge. Because of the trapping path formed by segments $ob$, $ba$, $ac$ and $co$, retro-transmission is not allowed not only along the waveguide, but also along any other propagation path.

\begin{figure} [th]
	\includegraphics[scale=0.275]{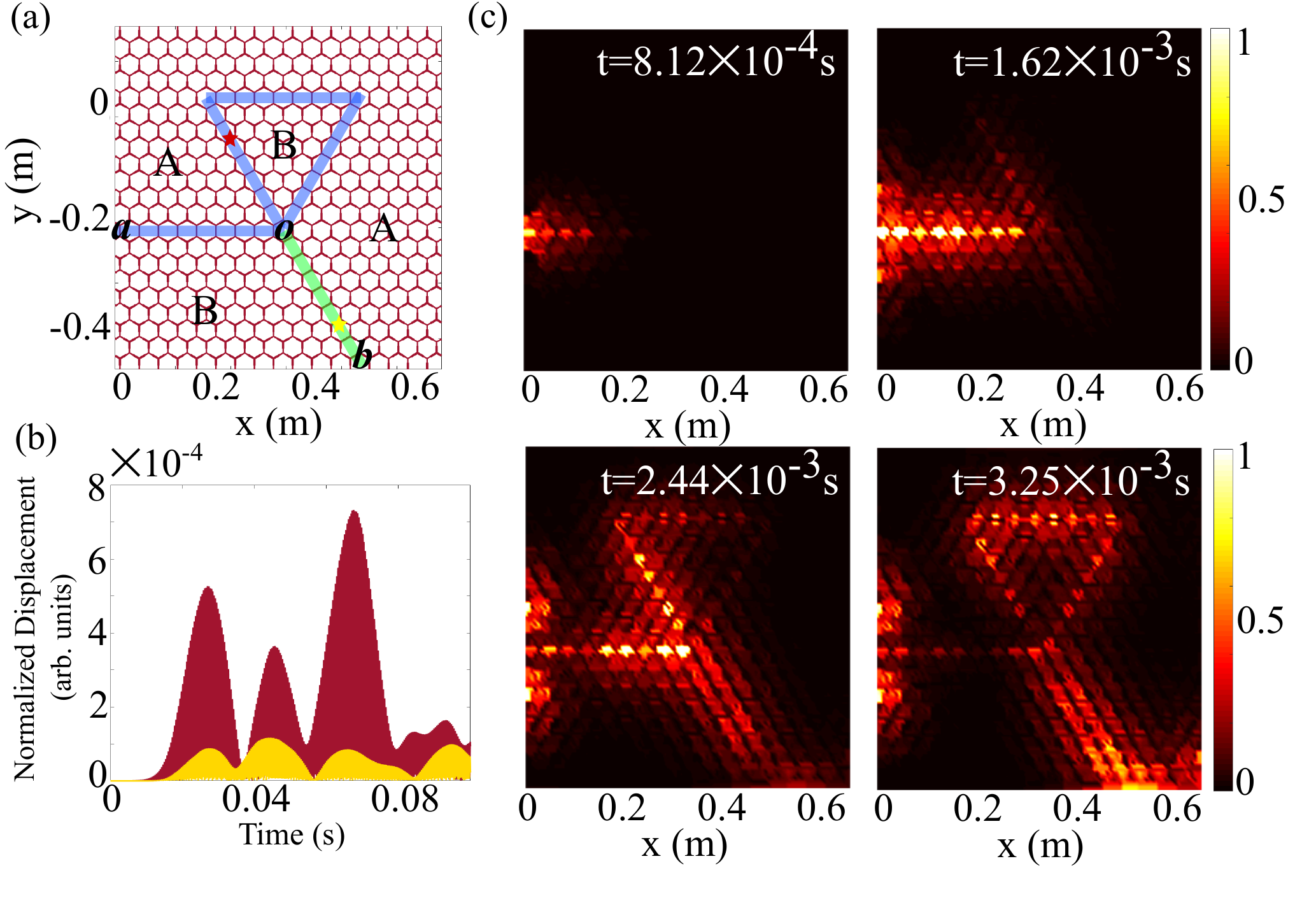} 
	\caption{(a) Full-scale simulation of a lattice with a triangular loop realized with topological interfaces. Thin and thick interfaces are marked in blue and green, respectively. (b) Magnitudes of displacements computed at the points marked by maroon and yellow stars along the thin and thick interfaces, respectively, in (a). (c) Snapshots of wavefield induced through a burst excitation at $\omega=8.707 \times 10^4$ Hz applied at port $a$. The displacement magnitudes are normalized by the largest value measured in the wavefield.}
	\label{loop}
\end{figure}

For our second example, we arrange three thin and one thick interfaces as shown in Fig.~\ref{loop} (a) to realize a triangular loop embedded in the lattice domain. The idea is to create an energy sink that locally traps the energy injected into the lattice from an edge and delays its release at another edge. When a wave traveling from port $a$ impinges onto junction $o$, the energy splits, with a portion propagating along the upper thin interface (which forms a $60^\circ$ angle with $oa$), and the other following the lower thick interface at a $120^\circ$ angle. The energy propagating upwards circulates clockwise in the triangular loop, eventually coming back to junction $o$. Here the energy splits again into a packet that enters the loop for a second cycle and a second packet that travels along the thick interface $ob$ and eventually reaches the lower edge after some delay, see Fig.~\ref{loop} (c). This process, which is repeated at every cycle, breaks down the burst of energy and releases it at port $b$ after finite intervals, with a period that depends on the geometric characteristics of the loop. As a result, in contrast with the input signal, which consists of a single packet, the output signal is broken down into multiple packets that are transmitted over an extended time window, as shown in Fig.~\ref{loop} (b).

The ability to create arbitrarily shaped and long waveguiding loops has implications for the design of efficient energy harvesting systems. Energy harvesting from mechanical vibrations requires the activation of transducers embedded in mechanical systems that convert mechanical energy into electrical energy to provide an electrical (e.g. voltage) output that can be harnessed and stored. Placing harvesters must be carried out compromising efficiency and parsimony: on one hand, one wants to maximize the output of the system by deploying many harvesters, on the other hand too many harvesters can be an impractical and expensive solution, in addition to detrimentally affecting the mechanical performance of the host system. Moreover, in a conventional medium subjected to a point excitation, the energy travels outwards from the excitation point with amplitude decreasing with the distance from the source. Therefore, even if harvesters were distributed densely in the domain, the output of those located far form the source would become progressively negligible. In contrast, one of the byproducts of waveguides is the ability to predict precisely where the energy from an excitation propagates. As a result, one could deploy harvesters along the waveguide path, thus making sure that they are all going to be engaged by the excitation and that the signal they experience overall maintains its amplitude (in the absence of significant damping and ignoring some inevitable dispersion-induced packet attenuation). In this respect, if the objective were to maximize the harvesting output, one would want to design long waveguides, whose total length may exceed the sample size, which would therefore require tortuous paths with corners and bends. The ability offered by QVHE to manage backscattering at these points and, more specifically, the possibility to create internal loops becomes then an invaluable asset.  It is important to note that, for this kind of application, a closed loop confined to the interior of the domain would be ideal, as it would force the energy to remain confined along the path (until the inevitable damping destroys the signal), thus allowing multiple harvesting events per transducer. The implementation of our junctions, however, requires the existence of an extra leg ($ob$), along which part of the signal is released in each period, and thus, from a strict internal energy harvesting perspective, lost.

\section{Conclusion}

In this work, we have experimentally demonstrated the in-plane dynamics of topologically-protected interface modes in honeycomb lattices with a bi-valued strut thickness pattern that breaks the space-inversion symmetry. We have documented similarities and differences between topological interfaces formed by connecting thin and thick portions of the unit cell. We have shown the robustness against back-scattering of wave propagation about $60^\circ$ corners with identical types of interfaces and about $120^\circ$ with mixed interfaces. Exploiting this knowledge, we have designed multi-interface junctions and used them to realize different structural logic circuits and elastic ports featuring unconventional wave manipulation effects at optical frequencies. Our results extend the applicability of valley-Hall effects to in-plane elastic dynamics and enrich the current understanding of these phenomena in mechanical systems by providing a new interpretation angle that blends topological and mechanistic arguments. The presented modular interface design philosophy can provide inspiration for a broad range of applications for energy management, trapping, routing and harvesting.  
\vspace{0.2cm}
\section{acknowledgment}
	The authors acknowledge the support of the National Science Foundation (NSF grant EFRI-1741618).

\nocite{apsrev41Control}
\bibliographystyle{apsrev4-2} 

%

\pagebreak
\widetext
\setcounter{equation}{0}
\setcounter{figure}{0}
\setcounter{table}{0}
\setcounter{page}{1}
\setcounter{section}{0}

\onecolumngrid
\begin{center}
\textbf{\large Supplemental Materials: Valley-Hall In-Plane Edge States as Building Blocks for Elastodynamic Logic Circuits}\\[.2cm]
  Jihong Ma,$^1$ Kai Sun,$^2$ and Stefano Gonella$^{1,*}$\\[.1cm]
{\itshape ${}^1$Department of Civil, Environmental, and Geo- Engineering, University of Minnesota, Minneapolis, MN 55455, USA\\
	${}^2$Department of Physics, University of Michigan, Ann Arbor, MI 48109, USA\\}
\end{center}
\twocolumngrid
\renewcommand{\thefigure}{S\arabic{figure}}
\renewcommand{\theequation}{S\arabic{equation}}
\renewcommand{\upperRomannumeral}[1]{\uppercase\expandafter{\romannumeral#1}}
\renewcommand{\lowerromannumeral}[1]{\romannumeral#1\relax}

	\section{\label{sec:app}Quantum valley hall effect theory}
	To determine whether a band-crossing at the $\boldsymbol{K}$ point for a lattice with $C_{6v}$ point group symmetry is a Dirac point, we calculate the Berry phase (or Berry flux) $\Phi$ around the $\boldsymbol{K}$ point:
	\begin{align}
	\Phi=\oint_v d\mathbf{k}\cdot \mathcal{\mathbf{A}}
	\label{Dirac}
	\end{align}
	where $v$ denotes a close loop around the $\boldsymbol{K}$ point and $\mathcal{\mathbf{A}}=-i \langle \boldsymbol{\psi}(\mathbf{k}) |\nabla_{\mathbf{k}}|\boldsymbol{\psi}(\mathbf{k})\rangle$ is the Berry connection, which can be obtained from the eigenvectors of a phonon band  $|\boldsymbol{\psi}(\mathbf{k})\rangle$. For a 2D system that is invariant under time-reversal and two-fold rotational (about $z$) symmetry, the Berry flux is generally quantized to $0$ or $\pi$ (up to mod $2\pi$), with $\Phi=\pi$ being the key signature of a Dirac point.
	
	Numerically, an efficient way to compute the Berry phase is via eigenvector overlap. Let us select $N$ wavevector points along a closed loop, labeled $i=1, \ldots, N$, which encapsulate the point of interest and such that the periodic boundary condition $\boldsymbol{\psi}_{N+1}=\boldsymbol{\psi}_{1}$ is satisfied. The inner product between eigenvectors at two neighboring points can be computed as $\langle \boldsymbol{\psi}_i | \boldsymbol{\psi}_{i+1}\rangle$. When $N$ is large and the wavevector points are densely populated along the loop, this inner product is in general a complex number with an absolute value close to unity. Multiplying the inner products computed between all the available pairs yields a complex number whose phase is the Berry phase:
	\begin{align}
	\Phi=\arg\left(\prod_{i=1}^{i=N} \langle \boldsymbol{\psi}_i | \boldsymbol{\psi}_{i+1}\rangle\right)
	\label{Dirac_cal}
	\end{align}
	As mentioned above, since our system preserves time-reversal and $C_{6v}$ point group symmetry, $\Phi$ is quantized to $0$ or $\pi$, and thus the product of eigenvectors appearing on the right-hand-side of Eqn. \eqref{Dirac_cal} is a real number. It is then sufficient to look at the sign of this product: if it is positive (negative), the corresponding Berry flux is $0$ ($\pi$), and thus a negative eigenvector overlap product is smoking-gun evidence for the existence of a Dirac point.
	
\begin{figure} [h]
		\centering
		\includegraphics[scale=0.18]{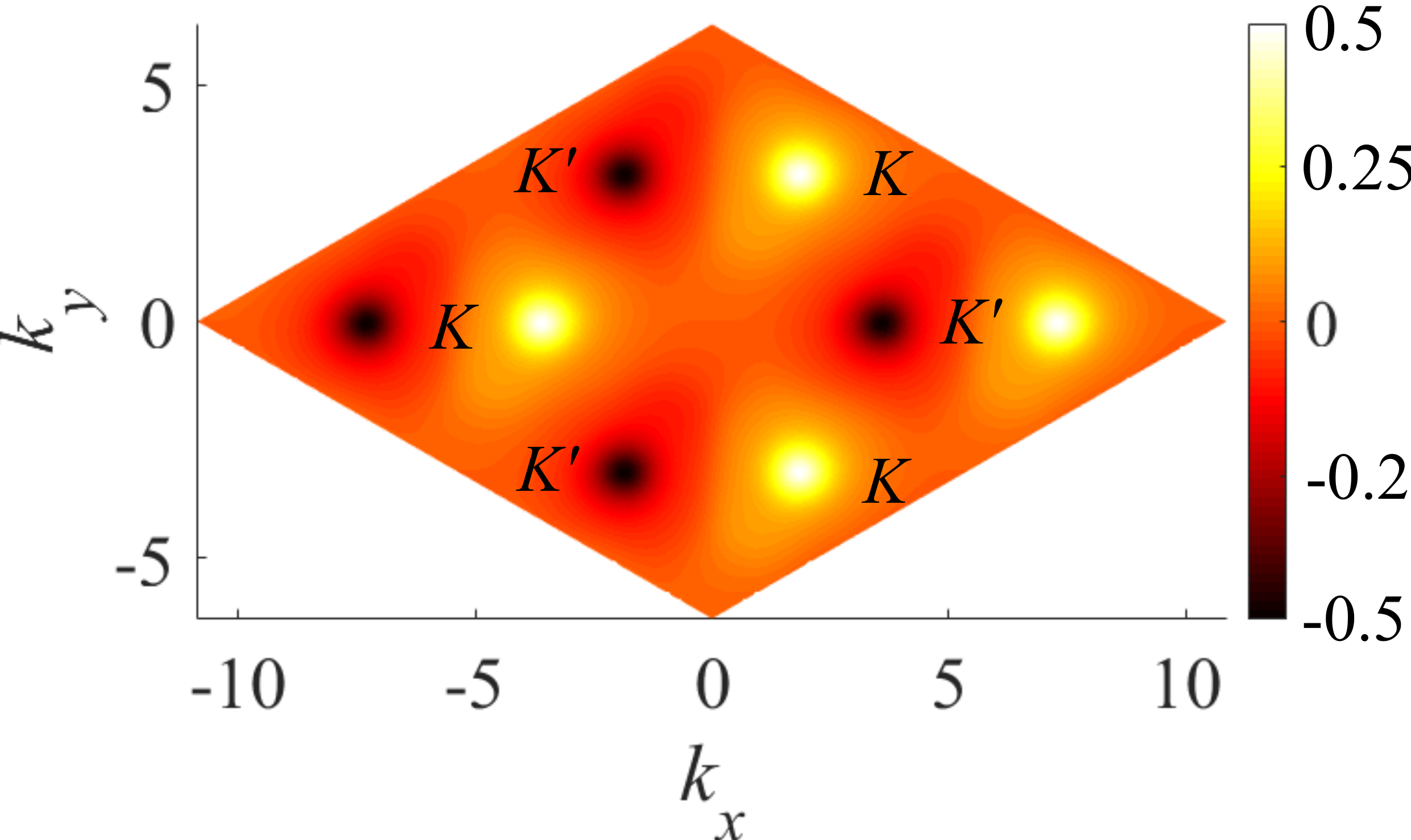} 
		\caption{Berry curvature computed for the $6^{th}$ band for the modulated hexagonal lattice shown in Fig. 1 (c) in the article showing peaks of amplitude approximately $\pm\frac{1}{2}$ at the valley points and featuring opposite signs between the $\textbf{\textit{K}}$ and $\textbf{\textit{K}}'$ points.}
		\label{BerryFig2}
	\end{figure}

	Once the space inversion symmetry is broken, the bandgap at the Dirac point is open. To describe the topology of this gap, we calculate the valley Chern number $C_{v}$:
	\begin{align}
	{\mathit{C_v}=\frac{1}{2\pi}\iint_v{B(\mathit{k_x,k_y})d\mathit{k_x}d\mathit{k_y}}},
	\label{Chern}
	\end{align}
	where $(\mathit{k_x,k_y})$ is an arbitrary point in the reciprocal space located in the neighborhood of the valley  $\boldsymbol{K}$ point, and $B(\mathit{k_x,k_y})$ is the Berry curvature calculated from the eigenvectors of the upper or lower bands near the band gap, which is expressed as:
	\begin{align}
	{\mathit{B(k_x,k_y)}=-2\cdot \mathrm{imag}\left<\frac{\partial \boldsymbol{\psi}}{\partial k_x}\right|\boldsymbol{M}\left|\frac{\partial\boldsymbol{\psi}}{\partial k_y}\right>},
	\label{Berry}
	\end{align}
	where $\boldsymbol{\psi}$ is the normalized eigenvector, and $\boldsymbol{M}$ is the unit cell mass matrix. A color map of the Berry curvature for the upper band is shown in Fig. \ref{BerryFig2}.  An analogous Berry curvature map, albeit with opposite signs, can be obtained for the lower band. The calculated Chern number at the valley $\boldsymbol{K}$ ($\boldsymbol{K'}$) should be +(-)0.5 according to classical quantum valley hall effect theory.
	
	\section{\label{sec:app}Toplogically-protected edge states}
\begin{figure} [h]
		\centering
		\includegraphics[scale=0.25]{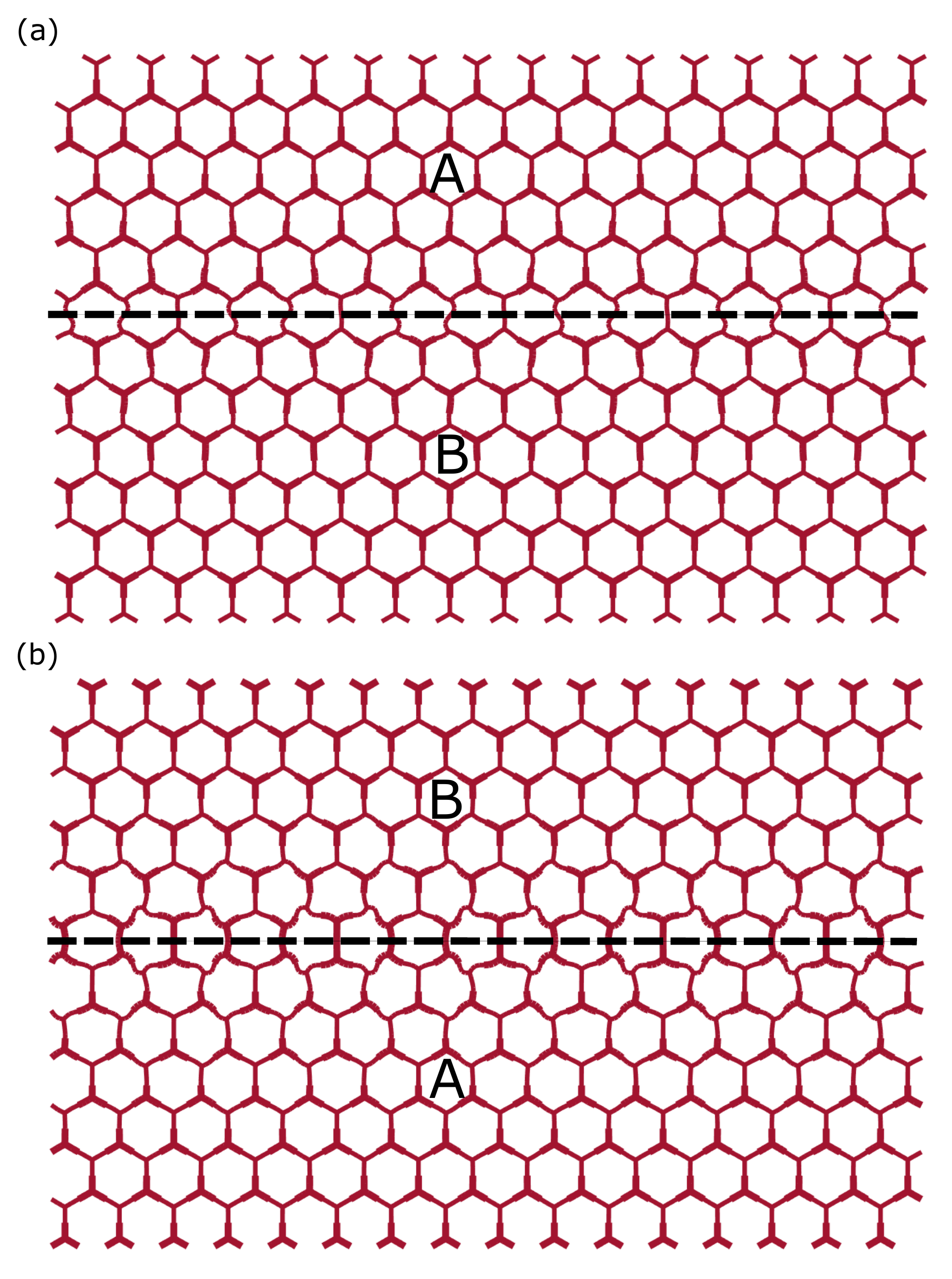} 
		\caption{Deformation patterns computed from the supercell under 1D Bloch conditions with (a) thin and (b) thick interfaces, revealing localized modes at the interfaces.}
		\label{SI_modeshape}
	\end{figure}

	The mode shapes calculated via finite element analysis (FEA) at the valley $\boldsymbol{K}$ point ($k_x=2\pi/3$) on the two emerging bands in the band gap in Fig. 2 (c) in the article are shown in Fig. \ref{SI_modeshape}. We can see that the two new modes are indeed interface modes as the deformation is mainly localized at the interface.
	
	One important feature of a topoligically protected edge states (TPES) is that it allows waves to propagate along sharp corners without appreciable back-scattering. Here we attempt to shed further light on the genesis of this phenomenon by providing an intuitive mechanistic explanation. To this end, we need to compute and plot the eigenvector phase rotation for a TPES. Let us note that TPES is, for all intents and purposes, a propagating wave concentrated near the interface boundary. Like any traveling wave, its displacement field is fully characterized in terms of an amplitude and a phase, i.e., 
	\begin{align}
	u(\mathbf{r},t)=A(\mathbf{r}) e^{i \mathbf{k}\cdot \mathbf{r}-i \omega t}
	\end{align}
	where the complex amplitude term $A(\mathbf{r})$ is the Bloch wave function and $\mathbf{k}\cdot \mathbf{r}$ is the phase, with $\mathbf{r}$ and $\mathbf{k}$ being the real space coordinate and the wavevector, respectively. For a TPES, the amplitude function is concentrated along the interface and decreases exponentially as we move away from the interface. The phase pattern of a TPES contains important information which directly dictates the wave propagation behavior at corners or junctions of interface boundaries.
	
	\begin{figure} [t]
		\includegraphics[scale=0.2]{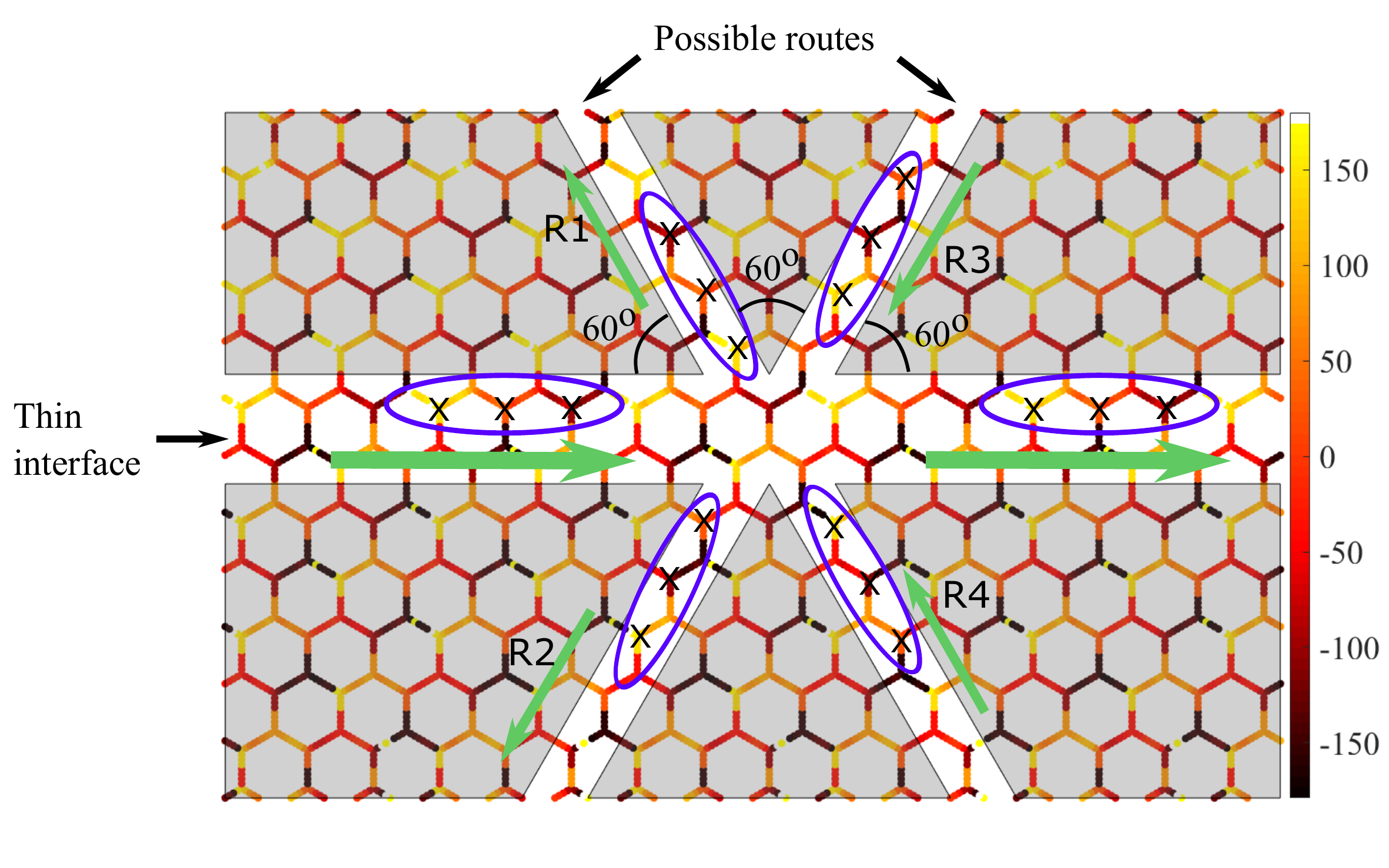} 
		\caption{Map of phase angles for the lattices with thin interface at $k_x=2\pi/3$. The color bar denotes the phase angle in $^\circ$. The propagation routes R1 to R4 denote possible zigzag interface paths that can be realized in the lattice. The green arrows denote the group velocity directions. The nodes marked with "X" are the reference nodes compared to determine compatibility (or lack thereof) between pairs of directions.}
		\label{phase}
	\end{figure}
	
	In Fig.~\ref{phase}, we plot the phase of displacement $u(\mathbf{r},t)$ (when t=0) for each node in the FEA model and we monitor the values at the nodes located along the interface defined in Fig. 2 (a) in the main article. Here, we focus our attention on a TPES with wavevector  $k_x =2\pi/3$, i.e., the $\boldsymbol{K}$ valley. As previously shown in Fig. 2 (c) in the article, this mode has a positive group velocity and thus corresponds to a mode that transports energy from left to right, as denoted by the green arrow in Fig.~\ref{phase}. From the colormap in Fig.~\ref{phase}, as can be seen, along the horizontal interface, the phase angle returns to the same value after every three unit cells following a high-zero-low sequence, which we term ``clockwise" (the opposite low-zero-high sequence will be called ``counter-clockwise"). The 3-cell periodicity and clockwise pattern are obtained for the selected wavenumber $k_x=+2\pi/3$, while the TPES at $k_x=-2\pi/3$ (not shown) has the same periodicity but counter-clockwise pattern.  Now, let us consider other lattice directions along which possible domain interfaces can be realized. Along R1 and R2, which form a $60^\circ$ angle with the horizontal interface, we observe that the phase also propagates clockwise. In other words, the phase pattern for a right-propagating TPES is fully consistent with TPES modes propagating along R1 and R2. 
	''maximize phase consistency, i.e., a wave will usually choose the propagation direction such that the in-coming and out-going waves have the highest phase coherence and therefore the most efficient energy transfer. For example, in a 3D homogeneous medium, the phase pattern of a plane wave is best matched by itself, and thus a wave will choose to maintain its wave-vector such that perfect phase match is preserved. However, if the wave impinges on a 2D interface, because phase coherence cannot be maintained in the whole 3D space anymore, the next best option is selected, i.e., to maintain the phase pattern along the 2D interface, leading to the mechanisms behind the well-known reflection and refraction patterns. 
	
	We can apply the same principle to the TPES. First, we consider a multi-segment interface which switches direction from the horizontal to either R1 or R2. At a corner, a wave packet propagating from left to right along the $x$-axis has two options: (a) being reflected towards the left or (b) propagation along R1 or R2. The phase pattern is fully preserved (clockwise in this case) for option (b), while in option (a), the phase will have to change its pattern to counter-clockwise, failing to match that of the incoming wave. Therefore, the wave will prefer traveling around the $60^\circ$ sharp corner, rather than being scattered back, which explains why a TPES can transmit through such a corner with nearly zero back-scattering.
	
	Let us consider now the propagation path that switches from the horizontal interface to R3 or R4, which form a $120^\circ$ angle with the horizontal direction, Note that for a 120 corner, the segment of R3 or R4 must involve the other interface type, otherwise the lattice cannot be properly completed. In this case, the landscape of phase rotation is reversed. Naively, we could conclude that R3 and R4 are unfavorable directions.  However, it is important to note that the TPES along this new interface features a negative group velocity [see the black curve in Fig. 2 (c) in the main article], opposite to that along the horizontal interface. Thus, for a wave packet propagating along R3 or R4, the phase pattern is also clockwise along the group velocity direction, consistent with the scenario of the incoming wave. As a result, waves can still propagate along such a $120^\circ$ angle, rather than being scattered back, provided that we switch to a different type of interface.

\end{document}